\documentstyle[aps,prb,epsf,multicol]{revtex}

\newcommand{\cntr}[1]{{\vspace*{\fill}\hspace*{\fill} #1
  \hspace*{\fill}\vspace*{\fill}}}

\begin{document}

\author{Xiao-jun Li and M. Schick\\
        Department of Physics, Box 351560 \\
        University of Washington, Seattle 98195-1560}
\title{Fluctuations in mixtures of lamellar- and nonlamellar-forming lipids}

\date{\today}
\maketitle
\begin{abstract}
We consider the role of nonlamellar-forming lipids in biological
membranes by examining fluctuations, within the random phase
approximation, of a model mixture of two lipids, one of which forms
lamellar phases while the other forms inverted hexagonal phases.  To
determine the extent to which nonlamellar-forming lipids facilitiate the
formation of nonlamellar structures in lipid mixtures, we examine the
fluctuation modes and various correlation functions in the lamellar
phase of the mixture. To highlight the role fluctuations can play, we
focus on the lamellar phase near its limit of stability. Our results
indicate that in the initial stages of the transition, undulations
appear in the lamellae occupied by the tails, and that the
nonlamellar-forming lipid dominates these undulations. The lamellae
occupied by the head groups pinch off to make the tubes of the hexagonal
phase. Examination of different correlations and susceptibilities makes
quantitative the dominant role of the nonlamellar-forming lipids.
\end{abstract}
\section{Introduction}
It is well known that the biological, lipid bilayer membrane contains
a mixture of lipids some of which, alone in aqueous solution, 
do not form bilayers at all.
This presents the interesting question as to what possible function
in the membrane these lipids 
can be serving\cite{prc,dekruijff,epand}. 
There are two lines of thought on the
answer.  The first is that the nonlamellar-forming lipids, roughly
characterized as having a small head group and long, splayed tails, can
with these tails increase significantly the local lateral pressure distribution in the 
interior of the bilayer. This increased pressure  may be necessary for a
protein or peptide to function\cite{swhui,sarah,bezrukov}. 
The second is that the nonlamellar-forming 
lipids can make less expensive the energetic cost to form a
small, nonlamellar region which may be needed in processes such as
membrane fusion\cite{siegel1,chern}. 
Experiment shows that there is the expected correlation
between the presence of nonlamellar-forming lipids and the ability
of the lamellar phase to make 
a transition to a nonlamellar, inverted-hexagonal
one\cite{tate,seddon}. 
There is also a correlation
between the concentration of such lipids and the ability to make other 
nonlamellar
structures, such as occur in membrane fusion\cite{sundler}.

In an attempt to shed some light on the role in membranes 
of nonlamellar-forming
lipids, we first introduced\cite{li1} a 
relatively simple and tractable model of
a lipid. We showed that it exhibited the 
polymorphism  observed in experiment as a function of a single
architectural parameter, the fraction of the total lipid volume occupied
by the head group. We also showed that the phase diagram of a model
aqueous lipid whose head group fraction was similar to that of 
dioleoylphosphatidylethanolamine was in good agreement with the measured
diagram\cite{dope,dope2}. We then considered a mixture of two lipids with the
same headgroup but different tail length\cite{li2}. Thus they
were distinguished by the different relative volume fractions of their
headgroups such that one formed lamellar, $L_{\alpha}$, phases while the other
formed inverted-hexagonal, $H_{II}$, ones. 
We examined the density profiles of the
tail segments from the two different lipids in each of the two
phases. In particular we determined that, in the $H_{II}$ phase of the
mixture, the density of the tail segments of the nonlamellar-forming
lipid varied around the Brillouin zone, with
its density being largest in the direction between next-nearest-neighbor
cores, as expected\cite{gruner}. We were able to make quantitative this
expected variation, and found the relative difference in tail segment
densities to be on the order of a few percent. 

The calculations in Ref. \cite{li1} and \cite{li2} were carried out
within self-consistent field theory, (SCFT), and therefore ignored 
fluctuations.
In this paper, we consider the effect of Gaussian fluctuations 
about the SCFT solution. This enables us to determine the effect of
fluctuations, within Gaussian order, on the transition from the
$L_{\alpha}$ to the $H_{II}$ phase, and to observe  the initial stages
of the  path between them. In order to highlight the effect of
fluctuations, we examine the metastable lamellar phase near its spinodal
in a region of the phase diagram in which the $H_{II}$ phase is the
stable one. We find that undulations occur in the lamellae
occupied by the tails, and that the lamellae occupied by the head groups
pinch off to form the tubes of the $H_{II}$ phase. This is similar to
scenarios proposed earlier by
Hui et al.\cite{hsb} and Caffrey\cite{caffrey}
and which arose from their experimental observations. The hexagonal phase
that is initiated by the lowest energy fluctuation mode is characterized
by a lattice parameter which is within 1\% of the stable $H_{II}$ phase,
and has the same orientation: one in which the tubes are coplanar with
the lamellae from which they were formed. This coplanarity is in
agreement with experiment\cite{photo}. We observe that the
nonlamellar-forming lipids dominate the undulations in the tail region,
while the lamellar-forming lipids tend to fill in the regions between
the cylinders as they pinch off. To make quantitative the effect of the
nonlamellar-forming lipids, we examine three different measures of
their role in the transition; the correlation between the density
fluctuation and the order parameter, the ratio of the susceptibilities
of the two different lipids to fluctuations at the critical wavevector,
and the overlap of the order parameter fluctuations of each lipid with the
equilibrium $H_{II}$ structure. All these measures show the
nonlamellar-forming lipids to play  a significant 
role in bringing about the transition.

Having calculated the fluctuations in Gaussian approximation, we are
able to calculate all structure factors. We show here the order
parameter-order parameter structure factor, which is the one
most readily measured, as well as the density-density structure factor 
of either lipid.

In the next section we review the theory of Gaussian fluctuations about
an ordered phase\cite{shind,laradji,schmid} and extend it to our model
of lipid mixtures. Results are presented next, and we
conclude with a brief summary. 

\section{Theory}
We consider an anhydrous mixture of $n_1$ lipids of
type 1 and $n_2$ lipids of type 2 in a volume $V$. Below we shall choose their
architecture so that type 1 lipids form lamellar phases while type 2
lipids form $H_{II}$ phases. All lipids consist of the {\it same} head group
of volume $v_h$ and two equal-length tails. Thus we model 
mixtures of  lipids drawn from a homologous series, such as the
phosphatidylethanolamines studied by Seddon et al.\cite{scm}. The local
density of the headgroups of lipid $L=1,2$, measured in units of the 
density $1/v_h$, is
denoted $\Phi_{h,L}({\bf r})$. Each
tail of lipid $L$ consists of $N\alpha_L$ segments 
of volume $v_t$.
The local density of lipid
tails of lipid $L$, again measured in units of $1/v_h$, is denoted
$\Phi_{t,L}({\bf r})$. The local volume fraction of these tail segments
is $(2Nv_t/v_h)\Phi_{t,L}\equiv 
\gamma_t\Phi_{t,L}$. The sole
architectural parameter, $f_L\equiv v_h/(v_h+2N\alpha_Lv_t)=
1/(1+\alpha_L\gamma_t)$, which characterizes each lipid is the relative 
volume fraction of its headgroup.

The only interaction in the system is between the headgroups and the
tails. The
interaction energy $E$ is
\begin{equation}
\label{energy}
 \frac{1}{kT}E[\Phi_{h,1}+\Phi_{h,2},\Phi_{t,1}+\Phi_{t,2}]
=\frac{2N\chi}{v_h}\int 
[\Phi_{h,1}({\bf r})+\Phi_{h,2}({\bf r})][\Phi_{t,1}({\bf r})+
\Phi_{t,2}({\bf r})]d{\bf r},
\end{equation}
where 
$\chi$ is the strength of the interaction,  and $T$ is the 
temperature.

The  partition function of the system can, without approximation, 
be written\cite{li2,matsen},
\begin{equation}
\label{pf2} 
{\cal Z}=\prod_{L=1}^2\int{\cal D}\Phi_{h,L}{\cal D}W_{h,L}{\cal D}
\Phi_{t,L}{\cal D}W_{t,L}\exp[-\Omega/kT],
\end{equation}
where ${\cal D}$ denotes a functional integral, and 
the grand potential $\Omega$ is given by
\begin{eqnarray}
\label{omega}
\Omega&=&-{kT\over v_h}\sum_{L=1}^2\left\{z_L{\cal Q}_L[W_{h,L},W_{t,L}]
+\int d{\bf r}[W_{h,L}\Phi_{h,L}+W_{t,L}\Phi_{t,L}]\right\}\nonumber\\
&+&E[\Phi_{h,1}+\Phi_{h,2},\Phi_{t,1}+\Phi_{t,2}],
\end{eqnarray}
with ${\cal Q}_L[W_{h,L},W_{t,L}]$ the partition function of a single
lipid of type $L$ in the external fields $W_{h,L}$ and $W_{t,L}$.

The integrals over the densities $\Phi_{h,L}$ and $\Phi_{t,L}$ could be 
carried out as the densities appear only quadratically, but the integrals 
over the fields $W_{h,L}$ and $W_{t,L}$ can not be. Thus some approximate 
evaluation must be made. The SCFT consists in
replacing the exact free energy, $-kT\ln{\cal Z}$ by the extremum of 
$\Omega$. In addition, we extremize this potential
subject to an incompressibility  constraint that the sum of all head and
tail densities is unity everywhere. We denote the values of the 
$\Phi_{h,L}$, $W_{h,L}$, $\Phi_{t,L}$ and $W_{t,L}$
which extremize $\Omega$ subject to this constraint 
by lower case letters. They are obtained from the following 
set of self-consistent equations:  
\begin{eqnarray}
\label{head}
\phi_{h,L}({\bf r})&=&-z_L{\delta{\cal Q}_L\over\delta w_{h,L}({\bf
r})},\qquad L=1,2,\\
\label{tail}
\phi_{t,L}({\bf r})&=&-z_L{\delta{\cal Q}_L\over\delta w_{t,L}({\bf
r})},\qquad L=1,2,\\
w_{h,L}({\bf r})&=&2\chi N\sum_{L'}\phi_{t,L'}({\bf r})-\xi({\bf
r}),\qquad L=1,2,\\
\label{tailfield}
w_{t,L}({\bf r})&=&2\chi N\sum_{L'}\phi_{h,L'}({\bf r})-
\gamma_t\xi({\bf r}),\qquad L=1,2,\\
\label{incomp}
1&=&\sum_{L}\phi_{h,L}({\bf r})+\gamma_t\sum_L\phi_{t,L}({\bf r}),
\end{eqnarray}
where $\xi({\bf r})$ is the Lagrange multiplier used to enforce the
incompressibility constraint.
The free energy in this approximation, 
$\Omega_{scf}$, is obtained from Eq. \ref{omega} by replacing the
densities and fields by their self-consistent values.

To include fluctuations about this self-consistent field result, 
one decomposes the fields,
$W_{h,L}$ and $W_{t,L}$, and the densities, $\Phi_{h,L}$ and
$\Phi_{t,L}$, into their  
self-consistent field values plus a fluctuating part; $W_{h,L}=w_{h,L}+\delta
W_{h,L}$, etc. One then substitutes these decompositions into the free
energy, Eq. \ref{omega}. Terms linear in the deviations vanish by
definition of the self-consistent field approximation. We consider all
quadratic terms in this paper so that the fluctuations are accounted for
within Gaussian approximation. Thus the exact partition function of
Eq. \ref{pf2} is approximated by
\begin{equation}
\label{pf3}
{\cal Z}\approx\exp[-\Omega_{scf}/kT]\prod_{L=1}^2\int{\cal D}\delta \Phi_{h,L}
{\cal D}\delta W_{h,L}{\cal D}\delta \Phi_{t,L}{\cal D}
\delta W_{t,L}\exp[-\Omega^{(2)}/kT].
\end{equation}
The expression for $\Omega^{(2)}$ is most easily written in matrix form.
Introduce the column matrices
\begin{equation}
\delta W=\left(\begin{array}{c}
                \delta W_{h,1}\\
                \delta W_{t,1}\\
                \delta W_{h,2}\\
                \delta W_{t,2}
                \end{array}\right)
\end{equation}
and
\begin{equation}
\delta \Phi=\left(\begin{array}{c}
                  \delta\Phi_{h,1}\\
                  \delta\Phi_{t,1}\\
                  \delta\Phi_{h,2}\\
                  \delta\Phi_{t,2}  
                  \end{array}\right)
\end{equation}
and the square matrices
\begin{equation}
J=\left(\begin{array}{cccc}
         0&1&0&1\\
         0&0&0&0\\
         0&1&0&1\\
         0&0&0&0
        \end{array}\right)
\end{equation}
\begin{equation}
K=\left(\begin{array}{cccc}
        z_1C_{hh,1}&z_1C_{ht,1}&0&0\\
        z_1C_{th,1}&z_1C_{tt,1}&0&0\\
        0&0&z_2C_{hh,2}&z_2C_{ht,2}\\  
        0&0&z_2C_{th,2}&z_2C_{tt,2}
        \end{array}\right)
\end{equation}
where
\begin{equation}
\label{funcderiv}
     C_{\alpha\beta,L}({\bf r},{\bf r}^{\prime})\equiv
V{\delta^2{\cal Q}_L[W_{h,L},W_{t,L}]\over
\delta W_{\alpha,L}({\bf r})\delta W_{\beta,L}({\bf r}^{\prime})}
\qquad \alpha,\beta={\rm h\ or\ t}.
\end{equation}
In terms of these matrices, the second order correction, $\Omega^{(2)}$, to the
self-consistent free energy is
\begin{eqnarray}
{v_h\Omega^{(2)}\over VkT}&=&{2\chi N\over V}\int d{\bf r}\delta\Phi^T({\bf
r})J\delta
\Phi({\bf r})\nonumber\\
&-&{1\over V}\int d{\bf r}\delta W^T({\bf r})\delta\Phi({\bf
r})
-{1\over 2V^2}\int d{\bf r}d{\bf r}^{\prime}\delta W^T({\bf r})K({\bf r},{\bf
r}^{\prime})\delta W({\bf r}^\prime),
\end{eqnarray}
where the superscript $T$ denotes transpose. The Gaussian integrals in
Eq. \ref{pf3} over
the four fields $\delta W_{h,L},\ \delta W_{t,L}$, $L=1,2$, can now be
carried out. This will leave $\Omega^{(2)}$ a 
functional of the quadratic density 
fluctuations. These fluctuations are not integrated over  
in order to leave displayed their coefficients, which  
are the inverse of the desired
density-density correlation functions. 
It is the inclusion of the field fluctuations to 
Gaussian order for given density fluctuations which constitutes the 
random phase approximation. 

Integration over the field variables  in Eq. \ref{pf3}  yields
\begin{equation}
{\cal Z}\approx{\cal N}_0\exp[-\Omega_{scf}/kT]\prod_{L=1}^2\int
{\cal D}\delta\Phi_{h,L}{\cal D}\delta\Phi_{t,L}
\exp[-\Omega_{rpa}/kT],
\end{equation}
where the factor ${\cal N}_0$ is the reciprocal of the square root of 
a determinant which is of no interest
here, and
\begin{equation}
\Omega_{rpa}={kT\over 2v_hV}\int d{\bf r}d{\bf
r}^{\prime}\delta\Phi^T({\bf r})\left[4\chi NVJ\delta({\bf r}-{\bf
r}^{\prime})
+K^{-1}({\bf r},{\bf r}^{\prime})\right]\delta \Phi({\bf r}^{\prime}).
\end{equation}
In addition to the random phase approximation, we now impose the 
 constraint of incompressibility, which reduces the number
of fluctuating fields from four, $\delta\Phi_{h,L},\ \delta\Phi_{t,L}$,
$L=1,2$, to three. We choose these three to be natural order parameters.
Define
the total local volume fractions of each lipid
\begin{equation}
\label{op1}
\Phi_L\equiv\Phi_{h,L}+\gamma_t\Phi_{t,L}, \qquad L=1,2,
\end{equation}
and the local difference between head and tail densities of each lipid
\begin{equation}
\label{op23}
\Psi_L\equiv\Phi_{h,L}-{1\over\alpha_L}\Phi_{t,L}, \qquad L=1,2.
\end{equation}
Note that as defined, the integral of $\Psi_L$ over the whole system vanishes.
In terms of these densities
\begin{eqnarray}\Phi_{h,L}&=&f_L\Phi_L+(1-f_L)\Psi_L,\\
               \Phi_{t,L}&=&\alpha_Lf_L[\Phi_L-\Psi_L].
\end{eqnarray}
We choose the three independent fluctuating quantities to be 
$\delta\Psi_1$ and $\delta\Psi_2$,
the fluctuations in the difference in head and tail densities of each
lipid, and $\delta\Phi_1$,
the fluctuation of the total density of lipid 1.
Because of the incompressibility constraint, $\delta\Phi_1=-\delta\Phi_2$.
The reduction in variables is easily written in terms of the column
matrix  
\begin{equation}
\delta\Theta=\left(\begin{array}{c}
                   \delta\Phi_1\\
                   \delta\Psi_1\\
                   \delta\Psi_2\end{array}\right)
\end{equation}
and the $4\times 3$ matrix
\begin{equation}
U=\left(\begin{array}{ccc}
        f_1&1-f_1&0\\
        \alpha_1f_1&-\alpha_1f_1 &0\\
        -f_2&0&1-f_2\\
        -\alpha_2f_2&0&-\alpha_2f_2\end{array}\right)
\end{equation}         
so that 
\begin{equation}
\delta\Phi=U\delta\Theta.
\end{equation}
Then the correction, $\Omega_{rpa}$ to the self-consistent free energy
can be written
\begin{eqnarray}
\Omega_{rpa}&=&{kT\over 2v_hV}\int d{\bf r}d{\bf
r}^{\prime}\delta\Theta^T({\bf r})U^T\left[4\chi NJ\delta({\bf r}-{\bf
r}^{\prime})
+K^{-1}({\bf r},{\bf r}^{\prime})\right]U\delta\Theta({\bf r}^{\prime}),
\nonumber\\
\label{definition}
&\equiv&{kT\over 2v_hV}\int d{\bf r}d{\bf
r}^{\prime}\delta\Theta^T({\bf r})
K_{rpa}^{-1}({\bf r},{\bf r}^{\prime})\delta\Theta({\bf r}^{\prime}).
\end{eqnarray}

There remains the calculation of the partition function 
${\cal Q}_L[W_{h,L},W_{t,L}]$ of a single lipid in the external fields
$W_{h,L}$ and $W_{t,L}$. Before carrying this out, we must specify further the
way in which the lipid tails are modeled. They are treated as
being completely flexible, with radii of gyration
$R_{g,L}=(N_La^2/6)^{1/2}$ for each tail.
The statistical segment length is $a$. 
The configuration of the 
$l$'th lipid of type $L$ is described by a space curve ${\bf
r}_{l,L}(s)$ where $s$ ranges from 0 at one end of one tail, through
$s=\alpha_L/2$ at which the head is located, to $s=\alpha_L$, 
the end of the other tail.

Because the tails are completely flexible, one can define the propagator
$q_L({\bf r},s|{\bf r}^\prime)$ which gives the relative probability of
finding the zero'th segment of a tail of a type $L$ lipid 
at ${\bf r}^\prime$ and the segment $s$ of the same tail at
${\bf r}$:
\begin{equation}
 q_L({\bf r},s|{\bf r}^\prime)=\int_{{\bf r}_{l,L}(0)={\bf
r}^{\prime}}^{{\bf r}_{l,L}(s)={\bf r}}\tilde{\cal D}{\bf
r}_{l,L}(s)\exp\left[-\int_0^s dt\ W_{t,L}({\bf r}_{l,L}(t))\right]
 \qquad 0\leq s\leq \alpha_l/2. \label{defq}
\end{equation}
 In this expression, $\tilde{\cal D}{\bf r}_{l,L}(s)$ denotes a functional
integral over the possible configurations of the tail of the lipid of type
$L$ and in which, in addition to the Boltzmann weight, the path is
weighted by the factor ${\cal P}[{\bf r}_{l,L};0,s]$ with
\begin{equation}
{\cal P}[{\bf r},s_1,s_2]={\cal
N}\exp\left[-\frac{1}{8R_g^2}\int_{s_1}^{s_2}ds|\frac{d{\bf
r}(s)}{ds}|^2
\right],
\end{equation}
where ${\cal N}$ is an unimportant normalization constant and
$R_g\equiv(Na^2/6)^{1/2}$ is the radius of gyration of a tail of length
$N$. This propagator is of central importance, both within the SCFT and
the random phase approximation.
Because the tails are flexible and execute a random walk in the
presence of the potential $W_{t,L}$, the propagator can be obtained
from the solution of the diffusion
equation
\begin{equation}
{\partial q_L({\bf r},s|{\bf r}^\prime)\over \partial s}=
2R_g^2\nabla^2q_L({\bf r},s|{\bf r}^\prime)-W_{t,L}({\bf r})q_L({\bf r},s|{\bf
r}^\prime),
\end{equation}
subject to the initial condition
\begin{equation}
q_L({\bf r},0|{\bf r}^\prime)=\delta({\bf r}-{\bf r}^{\prime}).
\end{equation}
The integrated propagator, or endpoint distribution function,
\begin{equation}
\tilde{q}_L({\bf r},s)\equiv\int d{\bf r}^{\prime}q_L({\bf r},s|{\bf r}
^{\prime})
=\int d{\bf r}^{\prime}q_L({\bf r}^{\prime},s|{\bf r}),
\end{equation}
is also useful.

From the propagator, the partition function of a single lipid of type
$L$ is obtained via
\begin{equation}
{\cal Q}_L[W_{h,L},W_{t,L}]=\int d{\bf r}_1d{\bf r}_2d{\bf r}_3
q_L({\bf r}_3,\alpha_L/2|{\bf r}_2)\exp[-W_{h,L}({\bf r}_2)]
q_L({\bf r}_2,\alpha_L/2|{\bf r}_1).
\end{equation}
Functional derivatives, such as those required in
Eqs. \ref{head}, \ref{tail}, and \ref{funcderiv}, are
obtained from the above and
\begin{equation}
{\delta q_L({\bf r},s|{\bf r}^{\prime})\over\delta W_{t,L}({\bf
r}_1)}=-\int_0^sdt\ q_L({\bf r},s-t|{\bf r}_1)q_L({\bf r}_1,t|{\bf
r}^{\prime}).
\end{equation}
The second functional derivatives of the single lipid partition function
are expressed in terms of the propagator;
\begin{eqnarray}
C_{hh,L}({\bf r},{\bf r}^{\prime})&=&V\delta({\bf r}-{\bf
r}^{\prime})\tilde{q}_L({\bf r},\alpha_L/2)e^{-W_{h,L}({\bf r})}
\tilde{q}_L({\bf r},\alpha_L/2),\\
C_{ht,L}({\bf r},{\bf r}^{\prime})&=&2Ve^{-W_{h,L}({\bf r})}
\tilde{q}_L({\bf r},\alpha_L/2)\int_0^{\alpha_L/2}ds\ q_L({\bf
r},\alpha_L/2-s|{\bf r}^{\prime})\tilde{q}_L({\bf r}^{\prime},s),\\
C_{th,L}({\bf r}^{\prime},{\bf r})&=&C_{ht,L}({\bf r},{\bf
r}^{\prime}),\\
C_{tt,L}({\bf r},{\bf r}^{\prime})&=&2[F_L({\bf r},{\bf r}^{\prime})+
F_L({\bf r}^{\prime},{\bf r})+G_L({\bf r},{\bf r}^{\prime})],
\end{eqnarray}
where
\begin{eqnarray}
F_L({\bf r},{\bf r}^{\prime})&=&V\int_0^{\alpha_l/2}ds\int_0^s dt
\ \tilde{q}_L({\bf r},\alpha_L/2-s)q_L({\bf r},s-t|{\bf r}^{\prime})
\times\nonumber\\
&&\int d{\bf r}_2
q_L({\bf r}^{\prime},t|{\bf r}_2)e^{-W_{h,L}({\bf r}_2)}
\tilde{q}_L({\bf r}_2,\alpha_L/2),
\end{eqnarray}
and\begin{eqnarray}
G_L({\bf r},{\bf r}^{\prime})&=&V\int_0^{\alpha_L/2}ds\int_0^{\alpha_L/2} dt
\ \tilde{q}_L({\bf r},\alpha_L/2-s)\tilde{q}_L({\bf r}^{\prime},\alpha_L/2-t)
\times\nonumber\\
&&\int d{\bf r}_2
q_L({\bf r},s|{\bf r}_2)e^{-W_{h,L}({\bf r}_2)}
q_L({\bf r}_2,t|{\bf r}^{\prime}).\label{defg}
\end{eqnarray}
Although Eqs. \ref{defq}--\ref{defg} are exact for arbitrary fields 
$W_{h,L}({\bf r})$ and $W_{t,L}({\bf r})$, only the mean-field solutions
$w_{h,L}({\bf r})$ and $w_{t,L}({\bf r})$ are needed in the random phase 
approximation.

Correlation functions in real space can now be calculated. Because there
are three independent order parameters, defined in Eqs. \ref{op1} and
\ref{op23}, there are six independent correlation functions, 
$<\delta\Theta_i({\bf r})\delta\Theta_j({\bf r}^{\prime})>$, $i,j=1,2,3$,
where the brackets indicate an average using the partition function
evaluated within the random phase approximation. We will be
interested in correlation functions which are various linear
combinations of these, so we define a general fluctuation
\begin{eqnarray}
\label{defvector} 
\delta\theta_a({\bf r})&\equiv&a_1\delta\Phi_1({\bf r})+a_2\delta\Psi_1({\bf
r})+
a_3\delta\Psi_2({\bf r}),\nonumber\\
&=&(a_1\ a_2\ a_3)\left(\begin{array}{c}
\delta\Phi_1({\bf r})\\
\delta\Psi_1({\bf r})\\
\delta\Psi_2({\bf r})\end{array}\right)\nonumber\\
&\equiv&{\bf a}\cdot\delta\Theta({\bf r}),
\end{eqnarray}
and consider the correlation of two such general fluctuations 
${\bf a}^T\cdot<\delta\Theta({\bf r})\delta\Theta^T({\bf
r}^{\prime})>\cdot{\bf b}$.
When the ensemble average is
carried out, one obtains 
\begin{equation} 
{\bf a}^T\cdot<\delta\Theta({\bf r})\delta\Theta^T({\bf
r}^{\prime})>\cdot{\bf b}={v_h\over V}\ {\bf a}^T\cdot K_{rpa}({\bf
r},{\bf r}^{\prime})\cdot{\bf b},
\end{equation} 
where $K_{rpa}({\bf r}, {\bf
r}^{\prime})$ is 
the inverse of $K_{rpa}^{-1}({\bf r}, {\bf r}^{\prime})$ defined in
Eq. \ref{definition}.

Because the phases of interest to us are ordered, it is advantageous to
exploit the space-group symmetry and, following Shi et al.\cite{shind},  to 
expand all functions of position in terms of Bloch states;
\begin{equation}
\psi_{n{\bf k}}({\bf r})=e^{i{\bf k}\cdot{\bf {r}}}\sum_{\bf G}
u_{n{\bf k}}({\bf G})e^{i{\bf G}\cdot{\bf {r}}}.
\end{equation}
Here ${\bf k}$ is a wavevector in the first Brillouin zone,  $n$ is the
band index, the vectors
${\bf G}$ are the reciprocal lattice vectors of the ordered phase, 
and the wavefunctions $u_{n{\bf k}}({\bf G})$ satisfy the Schr{\"o}dinger
equation
\begin{equation}
2R_g^2({\bf k}+{\bf G})^2u_{n{\bf k}}({\bf G})+\sum_{{\bf G'}}
\hat{W}_t({\bf G}-{\bf
G}^{\prime})u_{n{\bf k}}({\bf G}^{\prime})=
\epsilon_{n{\bf k}}u_{n{\bf k}}({\bf G}).
\end{equation}
The potentials, $\hat W_t({\bf G})$, are just the coefficients of  the
expansion of the periodic field $w_t({\bf r})$, Eq. \ref{tailfield} 
\begin{equation}
w_{t,1}({\bf r})=w_{t,2}({\bf r})=\sum_{\bf G}e^{i{\bf G}\cdot{\bf r}}\hat
W_t({\bf G}).
\end{equation}
The expansion of functions of two spatial coordinates 
$K({\bf r},{\bf r}^{\prime})$, such as the correlation functions, takes
the form
\begin{equation}
K({\bf r},{\bf r}^{\prime})=\sum_{n{\bf k};n'{\bf k}^{\prime}}
\left[{\hat K}\right]_{n{\bf k},n'{\bf k}^{\prime}}\psi_{n{\bf k}}({\bf
r})\psi_{n'{\bf k}^{\prime}}^*({\bf r}^{\prime})
\end{equation}
 
Experimentally measured structure factors are Fourier transforms of the
correlation functions
\begin{eqnarray}
S_{ab}({\bf q})&\equiv&{\bf a}^T\ \cdot {1\over V^2}
\int d{\bf r}d{\bf r}^{\prime}e^{-i{\bf
q}\cdot({\bf r}-{\bf r}^{\prime})}<\delta\Theta({\bf r})\delta\Theta^T({\bf
r}^{\prime})>\ \cdot{\bf b}\nonumber\\
&=&{\bf a}^T\cdot {v_h\over V}\sum_{n,n'}\left[{\hat K}_{rpa}\right]_
{n{\bf q}-{\bf G};n'{\bf q}-{\bf G}}u_{n{\bf q}-{\bf G}}({\bf G})
u^*_{n'{\bf q}-{\bf G}}({\bf G})\cdot{\bf b},
\end{eqnarray}
where ${\bf q}-{\bf G}$ lies in the first Brillouin zone.

All calculations are carried out in the basis of Bloch
functions. Therefore the important propagators $q_L({\bf r},s|{\bf
r}^{\prime})$ must be expanded in them, and the procedure outlined above
in real-space is followed in the space of Bloch functions. This is
straightforward, but tedious.  Many of the details can be found in the
Appendix of \cite{shind}, and we shall not repeat them here. We turn,
instead, to our results.

\section{Results}

We have chosen to model two lipids with the same head group, but with
different length tails; lipid 1 characterized by $\alpha_1=1$ so that 
its tails are of length N, and lipid 2 characterized by $\alpha_2=1.5$
so that its tails are 
of length 1.5 N. Because the headgroups are identical, the ratio 
$\gamma_t\equiv 2Nv_t/v_h$ is the same for each lipid, and we have
chosen $\gamma_t=2.5$. With these parameters, the volume of the head group
relative to that of the entire lipid is, for lipid 1, $f_1\equiv
1/(1+\alpha_1\gamma_t)=0.2857$, while that of lipid 2 is
$f_2\equiv1/(1+\alpha_2\gamma_t)=0.2105$. For comparison, the relative
head group volume of dioleoylphosphatidylethanolamine calculated from
volumes given in the literature\cite{randf} is $f=0.254$. 
From our previous work\cite{li2}, we
know that anhydrous lipid 1 forms a lamellar phase, while anhydrous
lipid 2 forms an inverted hexagonal phase.

For orientation, we reproduce in Fig. 1 the
phase diagram of the anhydrous system of the two lipids calculated
earlier\cite{li2}. The temperature, $T^*$, is defined in terms of the
interaction strength $T^*\equiv 1/2\chi N_1$, and $\Theta$ is the volume
fraction of lipid 1, the lamellar-forming lipid. Although we know of no
phase diagrams for anhydrous mixtures of lamellar- and non
lamellar-forming lipids with the same headgroup but different length
tails, the system for which we have made our calculation, there are
results for mixtures of lamellar-forming phosphatidylcholine and
hexagonal-forming phosphatidylethanolamine. Our phase diagram is similar
to those observed in anhydrous mixtures of
dilinoleoylphosphatidylethanolamine and
palmitoyloleoylphosphatidylcholine\cite{boni}, and for mixtures of
dioleoylphosphatidylcholine and dioleoylphosphatidylethanolamine and
10\% water by weight\cite{per}. Each show a significant region of cubic
phase between the inverted hexagonal phase, which dominates at low
concentrations of the lamellar-forming lipid, and the lamellar phase,
which dominates at high concentration.  The dashed line in
Fig. 1 is the calculated spinodal of the lamellar
phase; that is, for volume fractions of the lamellar-forming lipid which
are to the left of (i.e. less than) this line, the lamellar phase is
absolutely unstable. To the right of this line, but in the regions in
which either the inverted hexagonal, $H_{II}$ or inverted gyroid,
$G_{II}$, is stable, the lamellar, $L_{\alpha}$, phase is metastable. At
high temperatures, the system is disordered, $D$. We have indicated on
this phase diagram the two points at which we have calculated structure
factors; very near the lamellar spinodal line at $T^*=0.0625$,
$\Theta=0.356$,
at which point the fluctuations will be large, and at $T^*=0.0500$,
$\Theta=0.469$, far from the spinodal, where fluctuations will be much
reduced.  At the former point, the stable phase is $H_{II}$, while at
the latter it is the $G_{II}$.

The structure factor most readily measured experimentally is the
Fourier transform of the order parameter-order parameter correlation
function $<\delta\Psi({\bf r})\delta\Psi({\bf r}^{\prime})>$, with 
$\delta\Psi({\bf r})\equiv\delta\Psi_1({\bf r})+\delta\Psi_2({\bf
r})$ the total order parameter.
This order parameter is proportional to the difference in the 
densities of headgroups and tails. The structure factor is obtained by
setting the vectors ${\bf a}={\bf b}$, defined in Eq. \ref{defvector},
to be $(0,1,1)$.  
It has been calculated at  $T^*=0.0625$,
$\Theta=0.356$, and
is shown in Fig. 2. The normal to the lamellar planes is in the $z$
direction. The wave vectors $k_z$ and $k_x$ are measured in units of
$2\pi/D$, where $D$ is the equilibrium lattice spacing of the lamellar
phase. As in the block copolymer system\cite{laradji}, there is a large
response about the positions of the Bragg peaks $(k_x,k_z)=(0,\pm 1)$,
$(0,\pm 2)$, and $(0,\pm 3)$.
In addition there are four satellite peaks at
$(\pm 0.846, \pm 1/2)$. 
As shown below, these fluctuation modes lead to
undulations in the planes of head-groups  causing them to pinch off,
and to form tubes. 
The array of tubes produced by these fluctuations 
has a centered rectangular symmetry, c2mm. Its appearance is close to
hexagonal, but the structure lacks  the $\pi/3$ rotational symmetry
elements. If a hexagonal structure were formed from the
wavevectors $(0,1)$ and $({\sqrt 3}/2,1/2)$, the ratio of the
hexagonal spacing, $D_h$, to the lamellar spacing $D$ would be $2/{\sqrt
3}\approx 1.15$. We find that the actual ratio of the lattice parameter
of the stable,
(non-aqueous), hexagonal phase to that of the metastable
lamellar phase at the same temperature and composition to be 1.16. 
Thus the fluctuations do seem to be 
responsible for driving the lamellar phase to an ordered phase in which
cylinders are oriented parallel to the lamellae, and with the correct
spacing. The complete hexagonal
symmetry, however, cannot be
obtained from Gaussian fluctuations alone. We note that because the
lamellae are isotropic in the $x-y$ plane, all structure factors are
isotropic in the $k_x-k_y$ plane.

For comparison, the same order parameter-order parameter structure factor, 
but now calculated at $T^*=0.05$,
$\Theta=0.469$, at which point the $G_{II}$ phase is stable, is shown in
Fig. 3. While the structure about the Bragg peaks remains, there are no 
satellite peaks here, and therefore no tendency to break the lamellae of
headgroups into cylinders. 

The density-density structure factor, the Fourier transform of
$<\delta\Phi_1({\bf r})\delta\Phi_1({\bf r}^{\prime})>$, with 
$\delta\Phi_1=-\delta\Phi_2$
the fluctuation in the total density of lipid 1, is obtained by
setting the vectors ${\bf a}={\bf b}$, defined in Eq. \ref{defvector},
to be $(1,0,0)$. As our two lipids are drawn from a homologous series and
differ only in the length of the tails, this structure factor would
not be easily measured. It is shown in Fig. 4 at  $T^*=0.0625$,
$\Theta=0.356$, and in Fig. 5 at $T^*=0.05$, $\Theta=0.469$. This
structure factor is similar to the order parameter-order parameter
structure factor. (One discernable difference is the value at zero
wavevector which, for the density-density structure factor, is
related to the osmotic compressibility. This small peak probably
reflects the fact that it is rather easy to replace one lipid by the
other because they have identical headgroups, and differ only by the
length of their tails.)
That the two different structure factors of Figs. 2 and 4 are so similar
means that one of the two lipids tends to form
cylinders, while the other must form their complement due to the
constraint of incompressibility. Which lipid does which can not be
determined from this correlation function, but it can from the cross
correlation functions, as we now show.

Consider the correlation between the density and order parameter
fluctuations, $<\delta\Phi_1\delta\Psi>$. Because the headgroups
of all lipids are identical, we expect that changes in the order
parameter reflect, for the most part, changes in the densities of the
tails.  If the non
lamellar-forming lipid 2 dominates the order parameter, 
then we
expect that a positive fluctuation in the total tail density should be
correlated with a positive change in the density of lipid 2. 
As the order parameter is defined as the density of heads minus
the density of tails, a positive fluctuation in the tail density corresponds
to a negative fluctuation in the order parameter. In addition, a positive
fluctuation in in the density of lipid 2 corresponds to a negative
value of $\delta\Phi_1=-\delta\Phi_2$. As a consequence of the two
minus signs, we expect $<\delta\Phi_1\delta\Psi>$ to be positive if the
fluctuation in the non lamellar-forming tails dominate the order
parameter. If they do not, it will be negative. A limit on this 
correlation can be obtained by noting that 
\begin{equation}
<(\delta\Phi_1\pm\delta\Psi)^2>\ \geq 0.
\end{equation} 
 From this it follows that if
\begin{equation} R_1\equiv {2<\delta\Phi_1\delta\Psi>\over
<(\delta\Phi_1)^2>+<(\delta\Psi)^2>},
\end{equation}
then
\begin{equation}
1\geq R_1\geq -1.
\end{equation} 
A value 1 of this ratio means that the density of lipid 2 is completely
correlated with the order parameter,  a value $-1$ that the density of
lipid 1 is completely correlated with the order parameter, and a zero
value that there is no correlation between the density of either lipid
and the order parameter.

We have evaluated this quantity utilizing only the least stable
fluctuation mode, {\em i.e.} that mode with non-zero wavevector whose
energy will vanish at the spinodal. At $T^*=0.0625$,
$\Theta=0.356$, near the lamellar spinodal, this wave vector is again
$k_z=1/2,\ k_x\approx {\sqrt 3}/2$ .  We find
$R_1\approx 0.27,$ which indicates that it
is the nonlamellar forming lipid whose density is correlated with the
order parameter, and that the magnitude of this correlation is
appreciable.

As a second measure of the importance of the nonlamellar forming lipid
in the transition, we  have considered the susceptibilities
of each lipid to a disturbance with a wavevector equal to that of the
least stable mode. At the spinodal, all responses at this wavevector
will diverge, in general, but the amplitudes at which they do so will
vary. Thus we have evaluated
\begin{equation}
R_2\equiv{<\delta\hat\Psi_2({\bf k})\delta\hat\Psi_2({\bf -k})>\over
<\delta\hat\Psi_1({\bf k})\delta\hat\Psi_1({\bf -k})>} 
\end{equation}
at the wavevector of the least stable mode, where $\delta\hat\Psi_L({\bf
k})$ is
the Fourier component of $\delta\Psi_L({\bf r})$. At the same $T^*$
and $\Theta$ near the lamellar spinodal, we find $R_2= 2.05$. Thus the
susceptibility of the nonlamellar forming lipid to the perturbation at
this wavelength is twice that of the lamellar forming lipid.

A third measure of the relative importance of the two lipids is obtained
by comparing the order parameter  fluctuations of the two lipids to the
order
parameter in the $H_{II}$ phase itself. We do this 
as follows. Consider the first star of wavevectors 
of the equilibrium hexagonal
phase which is the stable one at the temperature and composition at
which the lamellar phase is metastable. 
It consists of six wavevectors, ${\bf k}_i$, $i=1$ to 6,  of
magnitude  $k^{(h)}=4\pi/{\sqrt 3}D_h$,
where $D_h$ is the lattice parameter of the $H_{II}$ phase. 
The first four can be
written as $(k_x,k_z)= k^{(h)}(\pm {\sqrt 3}/2, \pm1/2)$, 
and the other two
as $(k_x,k_z)=k^{(h)}(0,\pm 1)$. As noted earlier, if $D_h/D$ were equal to
$2/{\sqrt 3}$, the first four wavevectors would lie on the edge of the
Brillouin zone of the lamellar phase, while the others would correspond
to reciprocal lattice vectors of the lamellar phase. 
We can determine the Fourier components with
these six wavevectors of the 
order parameter fluctuations $\delta\Psi_2$ and $\delta\Psi_1$. This
tells us how large is the overlap of these fluctuations with the equilibrium
hexagonal structure. We denote this Fourier component 
$\delta\hat\Psi_L({\bf k}_i)$, $L=1,2.$ 
In general, it is complex so we calculate $\delta\hat\Psi_L({\bf k}_i)+
\delta\hat\Psi_L(-{\bf k}_i)$, which is real. 
At $T^*=0.0625$, $\Theta=0.356$, we  find

\begin{eqnarray}
R_3\equiv{\delta\hat\Psi_2({\bf k}_i)+\delta\hat\Psi_2(-{\bf k}_i)\over
\delta\hat\Psi_1({\bf
k}_i)+\delta\hat\Psi_1(-{\bf k}_i)}&=& 1.43\qquad i=1\ {\rm to}\ 4 \nonumber\\
      &=& 1.47\qquad i=5,6.
\end{eqnarray}

If the fluctuations had hexagonal symmetry, the two values would be
equal. We see that the amplitudes of the hexagonal 
fluctuations of the non lamellar-forming lipid are almost 50\% larger
than those of the lamellar-forming lipid.
 
We  next display the way in which the fluctuations of the lamellar phase
near its
spinodal indicate the beginning of the path to the stable $H_{II}$
phase. We examine the effect of the fluctuation mode with lowest energy
on the order parameter. In the absence of fluctuations, the order
parameter, which is the difference in the total head and tail segment
density,
 is 
\[\phi_h^{(1)}-{\phi_t^{(1)}\over\alpha_1}+\phi_h^{(2)}-{\phi_t^{(2)}
\over\alpha_2}\]
We add the  contribution to this order parameter
from the real part of the 
fluctuation mode with the lowest
energy,  $Re\{\delta\Psi({\bf r})\}$. 
We  add it with a variable amplitude,
$\epsilon$ to better visualize its effect\cite{laradji}. 
Fig. 6(a) shows the order parameter in mean-field theory, with no
fluctuation contribution, $\epsilon=0.$ The point in the phase diagram is 
$T^*=0.0625$, $\Theta=0.356$. 
The center of the headgroup region is at $z=0$. The lamellae are in the $x-y$
plane, and the coordinates are measured in units of the lattice spacing,
$D$. The grey scale 
divides the values from $-0.6 $ to $0.6$ into ten shades. The darkest
values are most positive, and correspond to a dominance of head groups. 
Values in (a) range from $-0.214$ to $0.251$. In (b), (c) , and (d),
the fluctuation
contribution is turned on with amplitudes $\epsilon =0.06$, $0.12$, and
$0.18$ respectively. The extreme values in (d) are $-0.301$ and $0.572$.
One clearly sees the undulation of the tail region, and the pinching off
of the head group region, leading to an $H_{II}$ phase in which the
location of the tubes is coplanar with the lamellar phase.
Again, the lattice spacing of the stable $H_{II}$
is within 1\% of the spacing expected if there were a perfect epitaxy
between the phases. 
The mechanism we see indicated here by the fluctuations from the
lamellar phase is very similar to those proposed by
Hui et al.\cite{hsb} and Caffrey\cite{caffrey}. The coplanarity which
is a direct result of it has been
observed by Gruner et al. with oriented photoreceptor membranes\cite{photo}.
We also note the possible significance for this problem of an
observation made in
Ref. \cite{laradji}. Due to isotropy in the $x-y$ plane, the fluctuation
modes are infinitely degenerate in the $k_x-k_y$ plane, that is, all
fluctuation modes with the same magnitude $k_x^2+k_y^2$ have the same
energy of excitation. If  one particular direction is preferentially
excited, the mode leads to the formation of ordered cylinders as we have
seen above. But if several directions are excited simulataneously, then
the ripples in the corresponding directions will interfere with one
another. One pattern that can be obtained is a periodic array of holes,
in the lamellae, producing a perforated lameller phase\cite{hajduk,wang}.
These perforations are similar to pores in the
membrane\cite{netz,muller}, and to
stalks\cite{siegel1,markin,chernomordik}.
 
Finally, we display the effect of the fluctuation mode with the lowest
energy on the difference in the tail distribution of the two lipids. 
The tail distribution of lipid $L$ as calculated in the
self-consistent field theory, without fluctuations, is
$\phi_{t,L}({\bf r})$. We add the contribution to this
distribution from the fluctuation mode with the lowest energy, 
$\epsilon Re\{\delta\Phi_{t,L}({\bf r})\}$ where $\epsilon$ is, again, an amplitude
which we can adjust.
In order to compare the tail distributions of the two
lipids, we must take account of the fact that, because the
concentrations of the two lipids are unequal, they have different
average tail segment densities
\begin{equation}
\bar\phi_{t,L}\equiv{1\over V}\int d{\bf r}\ \phi_{t,L}({\bf r}).
\end{equation}
Therefore we utilize the quantity
\begin{equation}
P_{t,L}(\epsilon, {\bf r})\equiv {\phi_{t,L}({\bf r})
+\epsilon Re\{\delta\Phi_{t,L}({\bf
r})\}\over\bar\phi_{t,L}}
\end{equation}
and plot 
\begin{equation}
\Delta P_t(\epsilon,{\bf r})\equiv
P_{t,1}(\epsilon, {\bf r})-P_{t,2}(\epsilon, {\bf r})
\end{equation}
for various values of $\epsilon$. This shows us how the fluctuations
about the lamellar phase tend to spatially arrange the two different
lipids. Again we place ourselves at the
point in the phase diagram at which $T^*=0.0625$, $\Theta=0.356$.
Figure 7(a) shows this difference within the self-consistent field
theory alone, that is with $\epsilon=0$. 
The center of the headgroup region is at $z=0$. The grey scale is
such that darkest values correspond to positive values of $P_t$ at which
lipid 1 dominates, and lightest values correspond to negative values of
$P_t$ at which the non lamellar-forming lipid 2 dominates. Thus the
density of lipid 2 is largest at the center of the tail region, which is
expected as lipid 2 has the longer tails. 
The maximum and minimum values on the grey scale of Fig. 7 correspond to
$-0.16$ and $0.16$, divided into ten shades of grey. The maximum and
minimum values attained by $\Delta P_t(0,{\bf r})$ are $0.089$ and
$-0.109$. Thus the differences in local tail segment densities shown in
Fig. 6 are on the order of $10\%$ of the average tail segment densities.

In figs. 7(b), (c), and (d) we turn on the contribution to the density
difference of the tail segments from the lowest energy fluctuation mode,
and plot $\Delta P_t(0.06,{\bf r})$, $\Delta P_t(0.12,{\bf r})$, 
and $\Delta P_t(0.18,{\bf r})$.
The largest and smallest values attained by $\Delta P_t(0.18,{\bf r})$ 
are 0.160
and $-0.147.$ In this sequence, one clearly sees that in the fluctuations,
the non lamellar-forming lipid dominates the  undulations of the tail
region, particularly in the direction of the second neighbors, the
direction which is most difficult for the tails to fill. The tails of
the lamellar-forming lipid are initially relegated to filling in the
region between the cylinders which form  within a given lamellae. 

In sum, we have employed a model system of a non-aqueous mixture of
lamellar- and nonlamellar-forming lipids to examine the Gaussian
fluctuations of the lamellar phase near its spinodal.  We have
calculated the structure factor which would be  experimentally measurable were
the lameller phase sufficiently oriented. We have found
that the initial stage of the $L_{\alpha}$ to $H_{II}$ transition
involves the formation of undulations in the tail-rich lamellae,
undulations which are dominated by the nonlamellar-forming lipid. As we
have noted, the lattice constant of the ordered, stable, $H_{II}$ phase
is within 1\% of that which would be produced by the lowest energy
fluctuation mode, and the orientations are, of course, identical. There
is also a coplanarity between one of the principal direction of the
tubes in the $H_{\alpha}$ and the orientation of the $L_{\alpha}$ phase
which produced it, a coplanarity observed in experiment\cite{photo}. We
have used several measures to show that the nonlamellar-forming lipid
plays a dominant role in this transition in the mixture. Because we have
examined the fluctuations very near the spinodal, we are presumably
seeing the beginning of spinodal decomposition. Further from the
spinodal, and in particular, near the $L_{\alpha}$ phase boundary which
is generally far from the spinodal (see Fig. 1), it is to be expected
that the transition proceeds by a nucleation and growth mechanism, which
may well involve some form of lipid intermediate
structure\cite{verkleij,siegel86,siegel97}.  Presumably by making a
series of controlled quenches from the lamellar to the hexagonal phase,
one should observe the progression from the spinodal decomposition
mechanism of undulations to the nucleation and growth intermediates. 
Certainly both mechanisms seem to have been observed in the same
system\cite{hsb}.
In either form of the
transition, structures which are nonlamellar must be created.  As we
have shown and made quantitative, the nonlamellar forming lipids tend
to dominate the process of this kind which we have examined. This 
lends support to the
line of argument that their role in the biological membrane is to
facilitate the creation of such structures.

This work was supported in part by the National Science Foundation under
grant number DMR9876864.

\newpage

\begin{description}
\item[Figure 1] Phase diagram of an anhydrous mixture of  lipids 1 and 2
in the plane of temperature, $T^*$, and volume fraction, $\Theta$, of
lipid 1, the lamellar-forming lipid. Phases shown are inverted hexagonal,
$H_{II}$, inverted gyroid, $G_{II}$, lamellar, $L_{\alpha}$, and
disordered, $D$. The dashed line is the spinodal of the $L_{\alpha}$ phase.
\item[Figure 2] Order-parameter order-parameter structure factor shown
in the $k_z-k_x$ plane. The wavevectors are in units of $2\pi/D$, with
$D$ the lamellar spacing. The point in the phase diagram, $T^*=0.0625$,
$\Theta=0.356$, at which the factor is calculated, is near the
spinodal of the lamellar phase.
\item[Figure 3] Same structure factor as in Fig. 2, but at the point in the
phase diagram $T^*=0.05$, $\Theta=0.469$, far from the spinodal of the
lamellar phase.
\item[Figure 4] Density density structure factor calculated at 
$T^*=0.0625$, $\Theta=0.356$.
\item[Figure 5] Density density structure factor calculated at 
$T^*=0.05$, $\Theta=0.469$.
\item[Figure 6] The total order parameter is shown in the $x-z$ plane
with different amplitudes, $\epsilon$, of the contribution from the 
fluctuation mode
with the lowest energy; (a) $\epsilon=0.$, (b) $\epsilon=0.06,$ 
(c) $\epsilon=0.12$, (d) $\epsilon=0.18$. Light areas indicate
regions in which tail density is greatest, dark regions where head
density is largest.
\item[Figure 7] The difference in the tail segment densities arising
from lipid 1
and from lipid 2. The amplitude of the contribution from the fluctuation mode
with the lowest energy is (a) $\epsilon=0.$, (b) $\epsilon=0.06,$ 
(c) $\epsilon=0.12$, (d) $\epsilon=0.18$. Light areas indicate regions
in which tails of lipid 2, the nonlamellar former, dominate, dark
regions the predominance of segments from the tails of lipid 1. 
\end{description}
%%%%%%%%%%%%%%%%%%%%%%%%%%%%%%%%%%%%%%%%%%%%%%%%%%%%%%%%%%%%%%%%%%%%%%
\begin{multicols}{2}
\narrowtext
%\newpage
\begin{figure}
\epsfxsize = 3.375in
\cntr{\epsfbox{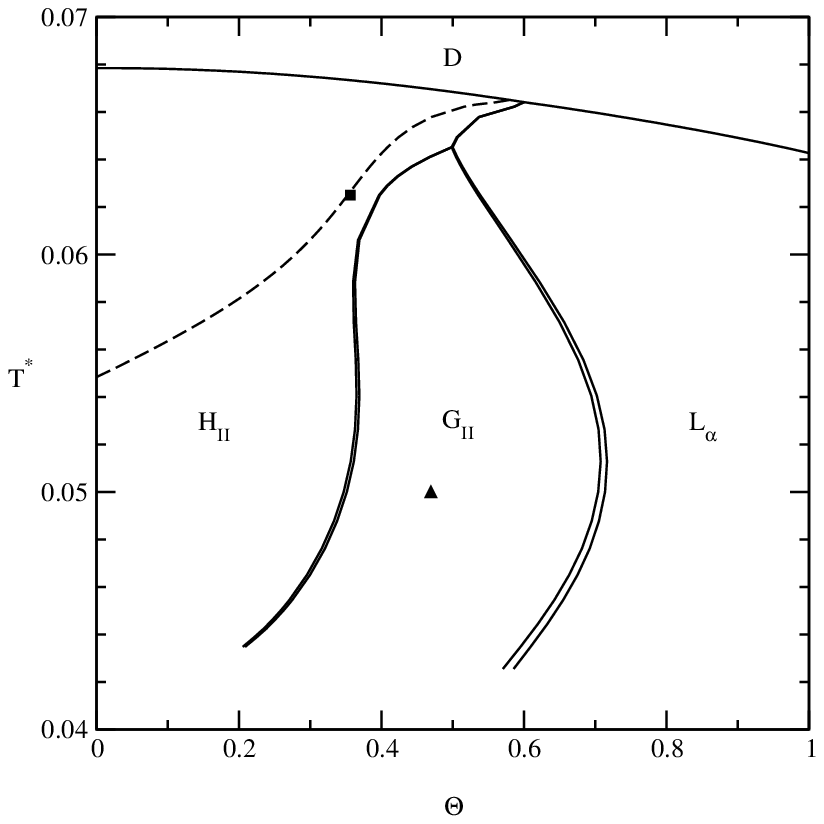}}
\caption{X.-j. Li, J. Chem. Phys.}
\label{phasediagram}
\end{figure}
%%%%%%%%%%%%%%%%%%%%%%%%%%%%%%%%%%%%%%%%%%%%%%%%%%%%%%%%%%%%%%%%%%%%%%
%\newpage
\begin{figure}
\epsfxsize = 3.375in
\cntr{\epsfbox{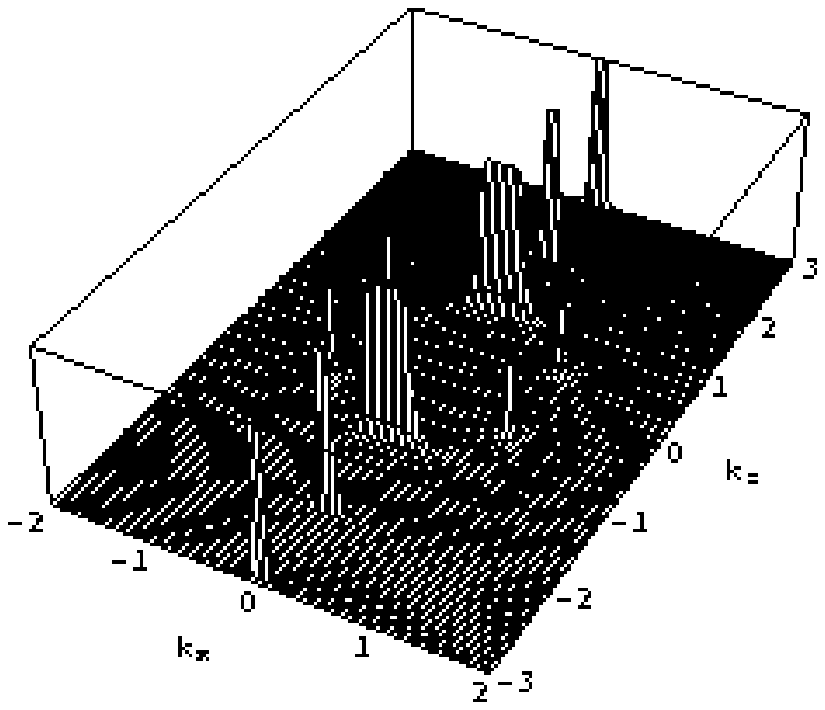}}
\caption{X.-j. Li, J. Chem. Phys.}
\label{psipsihex}
\end{figure}
%%%%%%%%%%%%%%%%%%%%%%%%%%%%%%%%%%%%%%%%%%%%%%%%%%%%%%%%%%%%%%%%%%%%%%
%\newpage
\begin{figure}
\epsfxsize = 3.375in
\cntr{\epsfbox{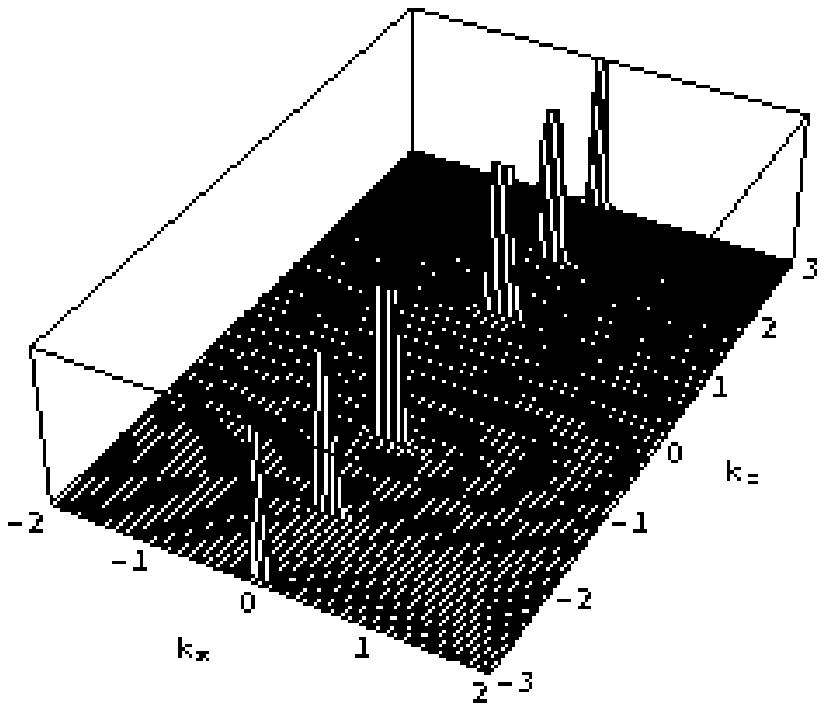}}
\caption{X.-j. Li, J. Chem. Phys.}
\label{psipsigyr}
\end{figure}
%%%%%%%%%%%%%%%%%%%%%%%%%%%%%%%%%%%%%%%%%%%%%%%%%%%%%%%%%%%%%%%%%%%%%%
%\newpage
\begin{figure}
\epsfxsize = 3.375in
\cntr{\epsfbox{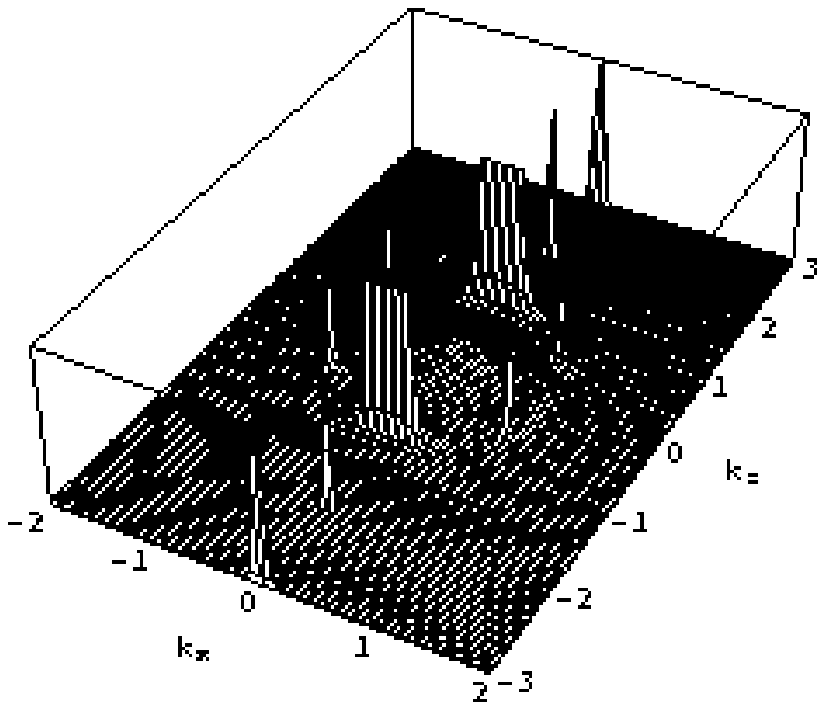}}
\caption{X.-j. Li, J. Chem. Phys.}
\label{sphiphihex}
\end{figure}
%%%%%%%%%%%%%%%%%%%%%%%%%%%%%%%%%%%%%%%%%%%%%%%%%%%%%%%%%%%%%%%%%%%%%%
%\newpage
\begin{figure}
\epsfxsize = 3.375in
\cntr{\epsfbox{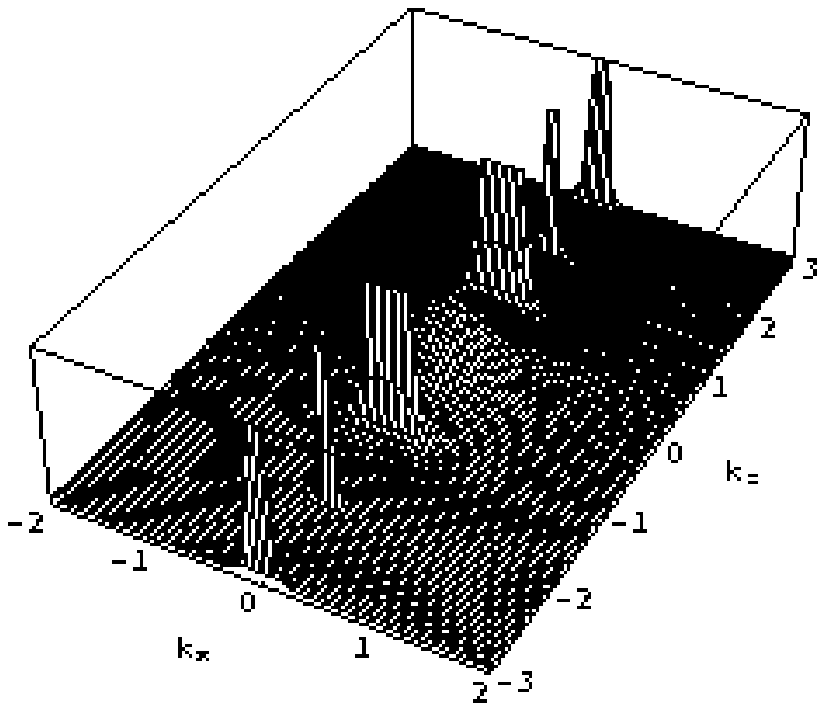}}
\caption{X.-j. Li, J. Chem. Phys.}
\label{sphiphigyr}
\end{figure}
%%%%%%%%%%%%%%%%%%%%%%%%%%%%%%%%%%%%%%%%%%%%%%%%%%%%%%%%%%%%%%%%%%%%%%
%\newpage
\begin{figure}
\epsfxsize = 3.375in
\cntr{\epsfbox{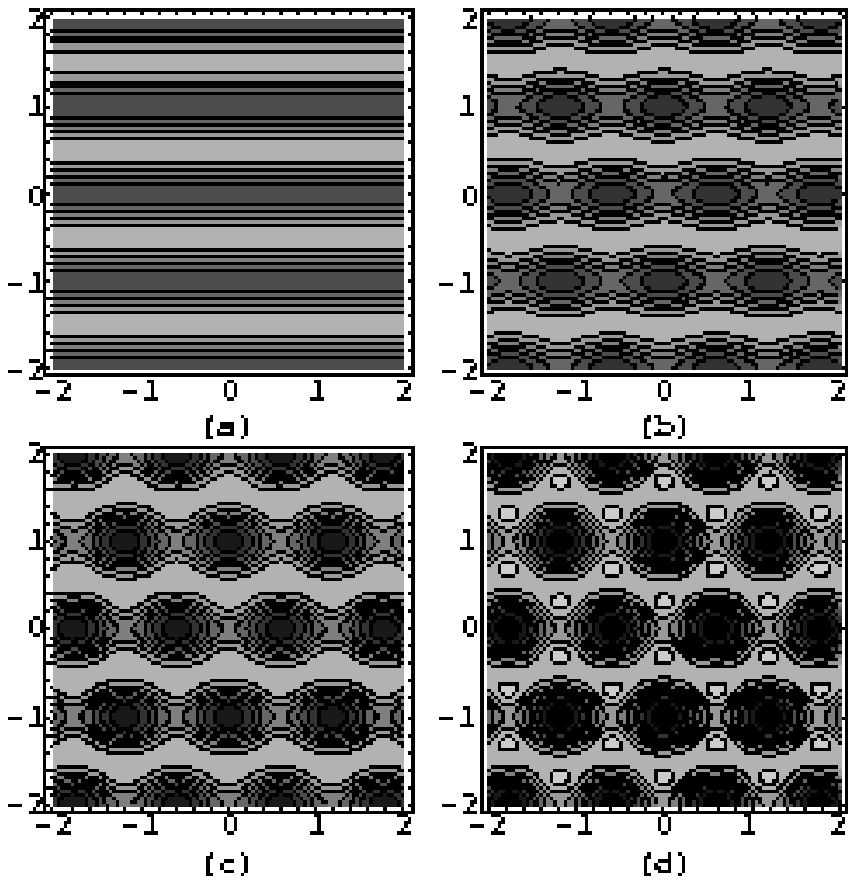}}
\caption{X.-j. Li, J. Chem. Phys.}
\label{deltaP1}
\end{figure}
%%%%%%%%%%%%%%%%%%%%%%%%%%%%%%%%%%%%%%%%%%%%%%%%%%%%%%%%%%%%%%%%%%%%%%
%\newpage
\begin{figure}
\epsfxsize = 3.375in
\cntr{\epsfbox{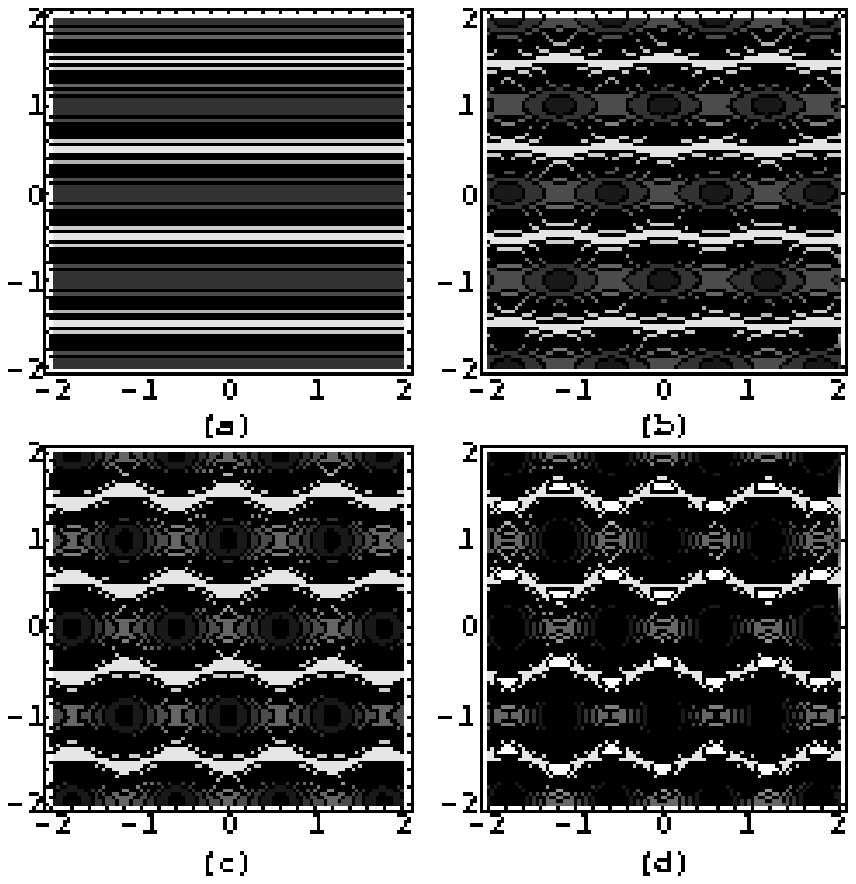}}
\caption{X.-j. Li, J. Chem. Phys.}
\label{deltaP2}
\end{figure}
%%%%%%%%%%%%%%%%%%%%%%%%%%%%%%%%%%%%%%%%%%%%%%%%%%%%%%%%%%%%%%%%%%%%%%
\end{multicols}
\end{document}